\documentstyle[12pt]{article}
\normalbaselines \oddsidemargin 0.5cm \evensidemargin 0.5cm
\begin{document}

\begin{center}
{\LARGE {\bf A classical cosmological model for triviality}}\\
\vspace{2cm} ${\bf Hadi~Salehi^{* \dag}},\vspace{1cm} {\bf P.~
Moyassari^{\dag}
}\footnote{e-mail:~P-Moyassari@cc.sbu.ac.ir.},\vspace{1cm} {\bf
R.~ Rashidi^\dag}$

\vspace{0.5cm} {\small {* Institute for Studies in Nonlinear
Analysis, School of Mathematical Sciences, \\ Shahid Beheshti
University, P.O.Box 19395-4716, Evin, Tehran 19834, Iran.}}\\
and\\ {\small {\dag Department of Physics, Shahid Beheshti
University, Evin,
Tehran 19839,  Iran.}}\\
\end{center}
\vspace{1cm} \vskip 1.5cm

\begin{abstract}
 The aim of this paper is to
study the triviality of $\lambda\phi^{4}$ theory in a classical
gravitational model. Starting from a conformal invariant scalar
tensor theory with a self-interaction term $\lambda\phi^{4}$, we
investigate the effect of a conformal symmetry breaking emerging
from the gravitational coupling of the large-scale distribution of
matter in the universe. Taking in this cosmological symmetry
breaking phase the infinite limit of the maximal length (the size
of the universe) and the zero limit of the minimal length (the
Planck length) implies triviality, i.e. a vanishing coupling
constant $\lambda$. It suggests that the activity of the
self-interaction term $\lambda\phi^{4}$ in the cosmological
context implies that the universe is finite and a minimal
fundamental length exists.
\end{abstract}

\vspace{1.5 cm}
\section{Introduction}
The simplest renormalized quantum field theory is $\lambda\phi^4 $
theory. There are strong evidences that $\lambda\phi^4 $ theory
is trivial \cite{1,2,3}. This feature is generally interpreted to
indicate that the interaction in quantum field theory must be
sensitive to some cutoff scale. Correspondingly the introduction
of a minimal fundamental length as a cutoff length is often
considered to be a categorical prerequisite to constructing an
interacting theory.

In the present paper, it is argued that triviality of
$\lambda\phi^4 $ theory can be interpreted as a cosmological
effect in a classical gravitational model. It indicates that the
concept of triviality can also appear at the classical level. It
is also argued that the avoidance of triviality necessiates the
introduction  of a maximal length in addition to a minimal length.

The paper is organized as follows: We consider a conformal
invariant scalar tensor theory with a self-interaction term
$\lambda\phi^4 $. A breakdown of conformal invariance is
investigated as the effect of the gravitational coupling of the
large-scale distribution of matter in the universe. In this
cosmological symmetry breaking phase the radius of the universe
and the Planck length play the role of the maximal length and the
minimal length respectively. We shall study two different limiting
procedures. In the first case we allow the maximal length to tend
to infinity and in the second case we let the minimal length to
tend to zero. It is then argued that both cases lead to
triviality. We also present a cosmological solution of the theory
to justify this result. According to this model the presence of
the self-interaction term $\lambda\phi^{4}$ in the gravitational
context implies a finite size for the universe and a non-vanishing
minimal length.

\section{The model}

We begin with the consideration of a conformal invariant
scalar-tensor theory based on the gravitational action
\begin{equation}
  S=-\frac{1}{2}\int d^4x\sqrt{-g}[g^{\alpha\beta}\nabla_{\alpha}\phi\nabla_{\beta}\phi+
 \frac{1}{6}R\phi^{2}-\frac{1}{2}\lambda\phi^{4}]\label{1b}
\end{equation}
where $\phi$ is a scalar field, $R$ is the scalar curvature
associated with the metric tensor $g_{\alpha\beta}$ and $\lambda$
is a dimensionless coupling parameter. Note that this action
differs from the standard conformal invariant gravitational action
\cite{4} by the presence of the self-interaction term
$\lambda\phi^4$.

The conformal invariance of the action (1) means that it is
invariant under a change in the local unit system, i.e. under
conformal transformations
\begin{equation}
\begin{array}{lll}
g_{\alpha\beta}\rightarrow\bar{g}_{\alpha\beta}=\Omega^2(x)g_{\alpha\beta}
\vspace{0.5cm} \\ \phi\rightarrow\bar{\phi}=\Omega^{-1}(x)\phi
\label{2b}
\end{array}
\end{equation}
in which $\Omega$ is a smooth dimensionless space-time function.
Therefore all conformal frames, i.e. all local unit systems, are
considered to be dynamically equivalent. In practice a particular
conformal frame may be singled out by a symmetry breaking effect
due to the gravitational coupling of a matter system. We
investigate the effect of cosmological symmetry breaking emerging
from the gravitational coupling of the large-scale distribution of
matter in the universe. In this cosmological symmetry breaking
phase the above action is generalized to the action of the
scalar-tensor theory
\begin{equation}
  S=-\frac{1}{2}\int d^4x\sqrt{-g}[g^{\alpha\beta}\nabla_{\alpha}\phi\nabla_{\beta}\phi+
 \frac{1}{6}R\phi^{2}+\mu^2\phi^2-\frac{1}{2}\lambda\phi^{4}]+S_{m}\label{1}
\end{equation}
here $S_m$ stands for the action of the large-scale matter in the
universe and $\mu$ is a parameter with the dimension of mass. The
parameter $\mu$ allows us to study breakdown of the conformal
invariance in this model. In general, under a conformal
transformation all dimensional parameters are required to be
transformed according to their dimensions so that $\mu$ should
obey transformation rule $\mu\rightarrow \Omega^{-1}\mu$. The
conformal invariance can, however, be broken when a particular
conformal frame is chosen in which the dimensional parameter $\mu$
takes on a constant configuration. Varying $S$ with respect to
$g^{\alpha\beta}$ and $\phi$ yields
\begin{equation}
    G_{\alpha\beta}-3\mu^2g_{\alpha\beta}+\frac{3}{2}\lambda\phi^2
    g_{\alpha\beta}=-
    6\phi^{-2}(T_{\alpha\beta}+
    \tau_{\alpha\beta})\label{2}
  \end{equation}
  \begin{equation}
  \Box\phi-\frac{1}{6}R\phi+\lambda\phi^3-\mu^2\phi=0  \label{3}
  \end{equation}
  where
   \begin{equation}
   T_{\alpha\beta}(g^{\alpha\beta})=\frac{-2}{\sqrt{-g}}\frac{\delta}{\delta g^{\alpha\beta}}
   S_{m}(
g^{\alpha\beta})\label{4}
\end{equation}
and
\begin{equation}
\tau_{\alpha\beta}=(\nabla_{\alpha}\phi\nabla_{\beta}\phi-
\frac{1}{2}g_{\alpha\beta}\nabla_{\rho}\phi\nabla^{\rho}\phi)
+\frac{1}{6}(g_{\alpha\beta}\Box-\nabla_{\alpha}\nabla_{\beta})\phi^{2}\label{5}
\end{equation}
here $T_{\alpha\beta}$ is the matter stress tensor of the
universe. Comparing the trace of (\ref{2}) with the equation (\ref{3}) yields:
\begin{equation}
g^{\alpha\beta}T_{\alpha\beta}=\mu^2\phi^2 \label{6}
\end{equation}
This relation shows that the breakdown of conformal symmetry is
related to the large scale distribution of matter in the universe
via a non-vanishing trace of $T_{\alpha\beta}$. In this case the
length scale $\mu^{-1}$ should be related to the typical size of
the universe $L$, namely $\mu^{-1}\sim L$. This condition is
characteristic to the cosmological symmetry breaking under
consideration.

It is possible from this condition to obtain an estimation
for the constant background average value of $\phi$ which
provides the strength of the gravitational coupling. In fact, the
trace of $T_{\alpha\beta}$ can be measured in terms of the
average density of the large scale distribution of matter, i.e.,
\begin{equation}
T^{\alpha}_{\alpha}=\mu^2\phi^2\sim M/L^{3}
\label{0-1}\end{equation}
which $M$ denotes the mass of
the universe. Now if one uses the empirical fact that the radius of the
universe coincides with its Schwarzschild radius $2GM$, one then
gets from (\ref{0-1})
an estimation of the constant background value of $\phi$, namely
\begin{equation}
\phi^{2}\sim\frac{M}{L} \sim G^{-1} \sim l_{p}^{-2}\label{1a}
\end{equation}
where $G$ and $l_{p}$ are the gravitational constant and the
Planck length respectively\footnote{In this note units are used in
which $\hbar=c=1$.}. Consequently the gravitational equations
(\ref{2}) reduce to the Einstein field equation with an effective
cosmological constant
\begin{equation}
\Lambda_{eff.}\sim
(\frac{-1}{L^{2}}+\frac{\lambda}{l_{p}^{2}}).\label{2a}
\end{equation}

This relation implies that in the symmetry breaking phase an
effective cosmological constant $\Lambda_{eff.}$ may be developed
as a consequence of two intrinsic cutoff scales, namely the
maximal length $L$ and the minimal length $l_p$. The former cutoff
contributes to $\Lambda_{eff.}$ through the the cosmological
symmetry breaking parameter $\mu\sim L^{-1}$ while the latter
cutoff contributes to $\Lambda_{eff.}$ through the coupling
constant of the self-interaction term $\lambda\phi^4$.

\section{Triviality}

We study the relation (\ref{2a}) with respect to two
different limiting procedures, namely the infinite limit of the
maximal cutoff and the zero limit of the minimal cutoff.

Let us consider the first case, in which the radius $L$ of the
universe is ideally taken to infinity. This limit has to be
carried out at constant physical conditions in the symmetry
breaking phase. Mathematically this means that the symmetry
breaking term $\mu^2 \phi^2$ is required to hold constant as
$L\rightarrow \infty $. By virtue of the relation (\ref{0-1}) the
last condition may alternatively be expressed by the requirement
that the limit $L\rightarrow \infty $ has to be carried out at
constant matter energy density $\sim \frac{M}{L^3}$. Thus we are
applying the infinite-volume limit or the thermodynamic limit of
large systems to the universe. Because of $\mu^{-1}\sim L$ this
limiting procedure requires through the relations
(\ref{0-1})-(\ref{1a}) that the gravitational constant goes to
zero. This means that it is interconnected with the the limit of a
vanishing minimal length $l_p$ through the cosmological symmetry
breaking effect. Thus both intrinsic cutoff scales drop out in
this limit.

In order to study the the implication of the limit $L\rightarrow
\infty$ for the coupling constant $\lambda$ we use the the
equation (\ref{2a}). As $L\rightarrow \infty$ the first term of
the right hand sight (\ref{2a}) drops out. Since the limit
$L\rightarrow \infty $ is interconnected with the limit
$l_P\rightarrow 0 $, the relation (\ref{2a}) demands that either
the coupling constant $\lambda$ must go to zero or we must deal
with an unacceptable magnification of the effective cosmological
constant. Since an infinite cosmological constant requires an
infinite vacuum energy density we discard the latter case. The
conclusion therefore is that the coupling constant $\lambda$ must
go to zero, that is the $\lambda \phi^4$ term disappears. Thus the
infinite limit of the maximal length leads to triviality
\footnote{This argument applies also if we simply assume that the
effective cosmological constant is zero as $L\rightarrow \infty$.
But this assumption is probably not correct if the infinite limit
is carried out at a constant matter energy density.}

Turning now to the second case. we ideally consider the zero limit
of the minimal length, that is the Planck length $l_{p}$ is taken
to zero. It follows from the relations (\ref{0-1})-(\ref{1a}) that
this limit is interconnected with the limit of the infinite radius
of the universe at constant matter energy density. Therefore
triviality in this case can be established by the same argument as
presented in the first case.

This consideration indicates that triviality can be related to two
limiting procedures, the infinite limit of the maximal length and
the zero limit of the minimal length. Both limiting procedures are
mutually connected through the relations (\ref{0-1})-(\ref{1a}).
It should be noted that this triviality argument was presented on
the basis of the relations  (\ref{0-1})-(\ref{1a}) which holds in
the conformal frame singled out by the cosmological symmetry
breaking effect we considered, namely the conformal frame in which
$\mu$ (or equivalently $\phi$) is constant.

\section{Cosmological solution}
In this section we attempt to obtain a cosmological solution for
the classical theory under consideration to justify the result of
the previous sections. We consider the homogeneous and isotropic
cosmological models in the interacting scalar tensor theory of
gravity (3). Accordingly, we start with the Robertson-Walker line
element and the energy tensor of a dust as the matter
\begin{equation}
\begin{array}{ll}  \vspace{1cm}
ds^2=dt^2-S^2(t)[\frac{dr^2}{1-kr^2}+r^2(d\theta^2+\sin^2\theta
d\varphi^2)] \\

 g^{\alpha\beta}T_{\alpha\beta}=\rho
\end{array}
\end{equation}
On account of the space time symmetry, we regard the scalar field
$\phi$ as a function of the cosmic time only. We introduce
 the quantities $H(t)$ and $F(t)$ through the relations
\begin{equation}
H=\dot{S}/S \hspace{1cm}
 F=\dot{\phi}/\phi
\end{equation}
where the overdot implies derivative with respect to $t$, cosmic
time. From the coefficient of the energy momentum tensor
$T_{\alpha\beta}$ in Eq.(4) one can postulate that the
 gravitational coupling $G$ behaves as
 \begin{equation}
 G\sim\phi^{-2}.
\end{equation}
 This assumption implies  $\mu^2=\rho G$, through the Eq.(8). With
 the above definitions, equations (4) and (5) lead to two independent
 equations

 \begin{equation}
\begin{array}{lll}
 H^2+k/S^2-\lambda/2G=\rho G+F^2+2HF \vspace{0.5cm} \\
 k/S^2-\dot{H}=3\rho G+F^2-\dot{F}+HF
 \end{array}
 \end{equation}\\

Now, we impose $k=1$ to assure a
 finite universe and proceed to present a solution to describe the
 universe in the present epoch. Thus, we consider the case $F=0$ or $G=$ constant which also
  implies the breakdown of conformal symmetry through choosing a constant
configuration for $\phi$. Also, to impose the characteristics of
the present epoch we approximate $H$ with $H_0$ in which $H_0$ is
the Hubble constant at the present epoch. Under these
considerations one can obtain the following constraint on
$\lambda$ and $\rho$

\begin{equation}
\lambda=4\rho G^2+2H_0^2G
\end{equation}

\begin{equation}
3\rho G=1/S^2
\end{equation}

In a closed universe, the radius of the universe is proportional
to the scale factor, $L\sim S$. Applying $G\sim l_{p}^2$ provides
the following relations
\begin{equation}
\mu\sim L^{-1}, \hspace{1cm} MG\sim L
\end{equation}
\begin{equation}
\Lambda_{eff}\sim-\frac{1}{L^2}+\frac{\lambda}{l^2_p}
\end{equation}
which are analogous to the equations (10) and (11). Therefore,
investigating this model in limit $l_p\rightarrow 0$ or
$L\rightarrow \infty$ at a fixed time, can establish the
triviality by  the same argument as presented in the third
section. On the other hand, it can be seen from equation (17) that
the exact dynamical solution of this theory determines the
configuration of $\lambda$ and the triviality also directly can be
obtained from this equation in the above limiting procedure. It
should be noted that, the limiting procedure at a fixed time
cannot be a dynamical limit, and is just an infinite volume limit
or thermodynamic limit of large systems.

\section{Concluding remarks}

The basic result of this paper is that the gravitational coupling
of the large-scale matter predicts a maximal cutoff length, namely
the size of the universe, to prevent the triviality of
$\lambda\phi^4$ theory in a classical gravitational model. This
means that the triviality of the self-interacting term
$\lambda\phi^4$ can be avoided by the cosmological characteristics
of a finite universe.

This is a result which has not appeared in the conventional
approach to triviality that has been established so far, where the
gravitational coupling of the large-scale matter is ignored and
only a minimal cutoff is suggested for the construction of an
interacting theory in quantum field theory. In fact, this
cosmological aspect of triviality indicates that the minimal
cutoff length and the maximal cutoff length may be considered as
being basically interconnected by the conditions of an interacting
theory in the gravitational context.

\end{document}